\documentclass[twocolumn]{aa}
\usepackage{graphicx}

\begin{document}

\title{Temperature determination via STJ optical spectroscopy}

\author{A.P. Reynolds, \inst{1}
        J.H.J. de Bruijne, \inst{1}
        M.A.C. Perryman, \inst{1}
        A. Peacock, \inst{1}
        C.M. Bridge \inst{2}
       }

\offprints{A.P. Reynolds}
\institute{Research and Scientific Support Department of ESA, ESTEC,
           Postbus 299, 2200 AG Noordwijk, The Netherlands
\and
           Mullard Space Science Laboratory, University College London,
           Holmbury St. Mary, Dorking, Surrey, RH5 6NT, UK}

\date{}

\authorrunning{A.P. Reynolds et al.}
\titlerunning{STJ-based temperature determination}

\abstract{ESA's Superconducting Tunnel Junction (STJ) optical
photon-counting camera (S-Cam2) incorporates an array of pixels with
intrinsic energy sensitivity. Using the spectral fitting technique
common in X-ray astronomy, we fit black bodies to nine stellar
spectra, ranging from cool flare stars to hot white dwarfs. The
measured temperatures are consistent with literature values 
at the expected level of
accuracy based on the predicted gain stability of the instrument.
Having also demonstrated that systematic effects due to count rate are
likely to be small, we then proceed to apply the temperature
determination method to four cataclysmic variable (CV) binary
systems. In three cases we measure the temperature of the accretion
stream, while in the fourth we measure the temperature of the white
dwarf. The results are discussed in the context of existing CV
results. We conclude by outlining the prospects for future versions of
S-Cam.
}

\maketitle

\section{Introduction}

The determination of astrophysical temperatures in the optical usually
requires photometric filtering systems or dispersive optics. Recent
developments in high-efficiency superconducting detectors, however,
have opened the possibility of measuring photon energies directly
(Perryman et al. 1993; Peacock et al. 1996). While the energy
resolution possible with this technology is presently modest, it has
already proven invaluable for studying faint, rapidly varying
astrophysical sources such as cataclysmic variables. The study of
colour ratio variations in these systems has already yielded
significant new insights into the accretion regimes, the logical next
step being to derive temperatures for the various system components.

ESA has developed two prototype cameras based on an array of STJ
devices, and high time-resolution spectrally-resolved observations of
rapidly variable sources such as cataclysmic variables and optical
pulsars have been reported (Perryman et al. 1999, Perryman et
al. 2001, Bridge et al. 2002a, Steeghs et al. 2002, Bridge et
al. 2002b). In addition, we have now demonstrated that we can measure
quasar redshifts directly with a precision of $\sim$ 1\% (de Bruijne
et al. 2002). The quasar study involved fitting a synthetic model to
the observed distribution of photon energies in a given observation,
and then adjusting the model redshift until a best-fit was
obtained. In this paper we perform a similar study on a small sample
of stellar observations. We demonstrate that we can fit black bodies
to the data and thereby extract reasonable estimates of $T_{\rm eff}$
for objects spanning a range of temperatures between $\sim$
3000---15\,000~K. Having demonstrated the validity of the method, we
then apply it to the specific case of accretion regions in several
cataclysmic variable binary systems.

\begin{table*}
\caption[]{Stellar targets for temperature determination: literature
values and S-Cam2 results. The column headed ND indicates the neutral
density filter setting. Upper and lower values for $T_{\rm eff}$ (Obs)
show temperature extremes corresponding to gain variations of
$\pm$1.0\%.}
\begin{center}
\begin{tabular}{rcrcllll}
\hline
\hline
\multicolumn{1}{c}{No} & \multicolumn{1}{c}{Date} & Exposure (s) & \multicolumn{1}{c}{ND} & Star & Type (Lit) & $T_{\rm eff}$ (Lit; K) & $T_{\rm eff}$ (Obs; K)\\
\hline\\[-9pt]
 1 & 1999-12-09 &  600 &  3  & AD Leo      & M3.5V       & $           3400$ & $3100^{3300}_{3000}$\\
 2 & 1999-12-10 & 1320 & --  & G117$-$B15A & DA          & $11\,500-12\,620$ & $13\,400^{16\,300}_{11\,200}$\\
 3 & 1999-12-16 &  900 &  2  & G191$-$B2B  & DAw         & $        56\,000$ & --\\
 4 & 1999-12-16 & 1200 &  3  & Feige 15    & A0          & $        10\,800$ & $10\,400^{12,800}_{9100}$\\
 5 & 1999-12-16 & 1800 &  2  & EV Lac      & M3.5        & $           3300$ & $3200^{3500}_{3100}$\\
 6 & 1999-12-16 &  900 &  2  & Feige 25    & B7          & $        12\,800$ & $11\,400^{14,100}_{9950}$\\
 7 & 2000-04-27 &  600 & --  & G138$-$31   & DA8         & $      6300-6870$ & $7200^{8100}_{6600}$\\
 8 & 2000-04-28 &  600 &  1  & HZ 21       & DA          & $        48\,000$ & $100\,000^{100\,000}_{40\,700}$\\
 9 & 2000-04-28 &  600 & --  & GD337       & DA$+\ldots$ & $   9250-15\,000$ & $11\,300^{13\,600}_{9900}$\\
10 & 2000-09-29 &  300 &  3  & BD$+$28 4211& Op          & $        82\,000$ & $44\,900^{100,000}_{26\,000}$\\
\hline
\hline
\end{tabular}\\
\end{center}
\begin{flushleft}
1: Legget et al. (1996);\\
2: Robinson et al. (1995) favour 12\,375 K; 
Bergeron et al. (1995) favour 12\,620~K; Fontaine et al. (1996) 
favour 11\,500~K; mean value (12\,165~K) plotted in Fig~2\\
3: Wolff et al. (1998), used for calibration purposes in this paper;\\
4: $T_{\rm eff}$ estimated from A0 spectral type (Zombeck 1990);\\
5: Amado \& Byrne (1997);\\
6: Keenan \& Dufton (1983);\\
7: Greenstein (1984) gives DA8, which implies $T_{\rm eff}$ = 6300~K 
(Sion et al. 1983); Bergeron et al (2001) give $T_{\rm eff}$ = 6870~K; mean
value (6585~K) plotted in Fig~2;\\
8: Koester et al. (1979);\\
9: $T_{\rm eff} = 9250$~K derived by authors using $G-I$ data and 
$G-I$ versus $T_{\rm eff}$ relationship in 
Greenstein (1976); $T_{\rm eff} = 15\,000$~K\hfill\break\null\hspace{0.35 
truecm} from $G-R$ colours in Greenstein (1975), 
using $G-R$ versus absolute magnitude ($M_{1.85}$) and $M_{1.85}$ 
versus $T_{\rm eff}$\hfill\break\null\hspace{0.35 truecm} 
calibrations in Greenstein (1976); mean value (12125~K)plotted in Fig~2;\\
10: Herbig (1999).\\
\end{flushleft}
\label{tab:observations}
\end{table*}

\section{Observations}

\subsection{The sample}

The sample of stellar observations consists of flux standards and
scientific targets observed during three campaigns on the 4.2--m
William Herschel Telescope on La Palma. Since these observations were
conducted for reasons other than this study, the data set is rather
inhomogeneous with regards to neutral density filter settings, seeing
conditions and S/N. The observations are listed in Table~1. Spectral
types are assigned on the basis of the current SIMBAD entry, while
$T_{\rm eff}$ values are derived from the literature. The hot white
dwarf G191-B2B was used to constrain the instrumental response at high
energies and is therefore omitted from the subsequent discussion.

\begin{figure}
\centering
\includegraphics[angle=-90,width=8.43truecm]{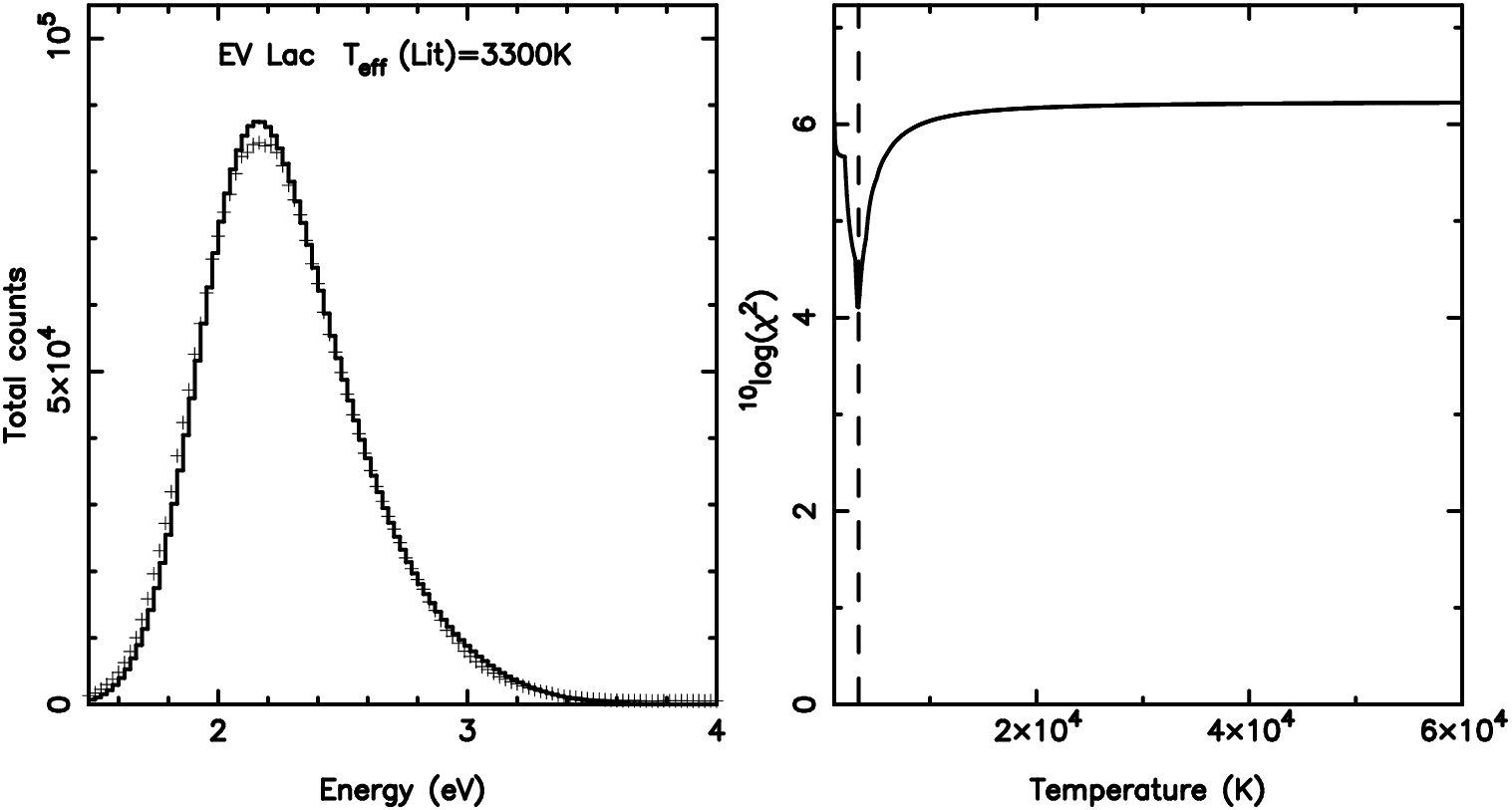}
\includegraphics[angle=-90,width=8.43truecm]{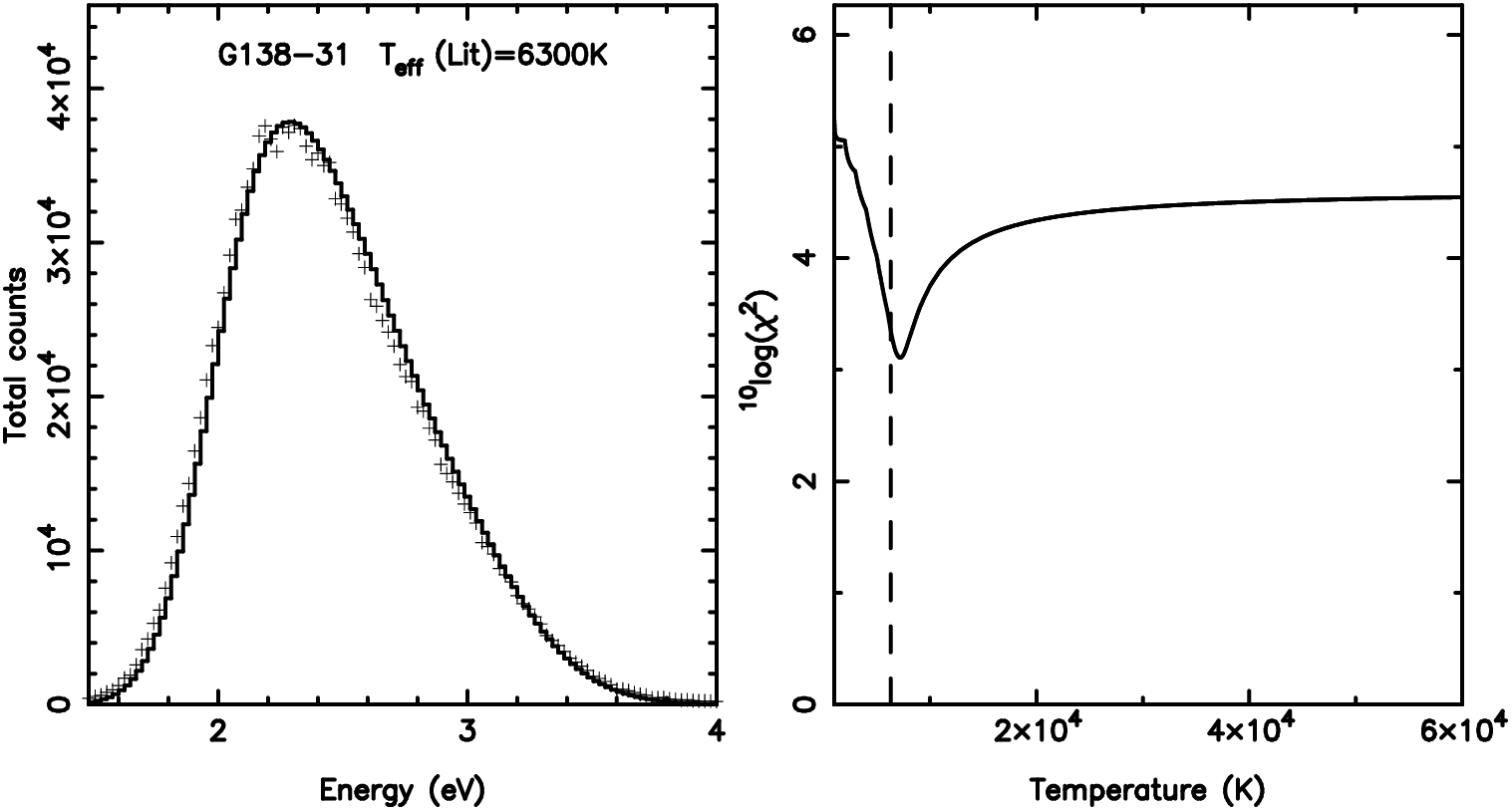}
\includegraphics[angle=-90,width=8.43truecm]{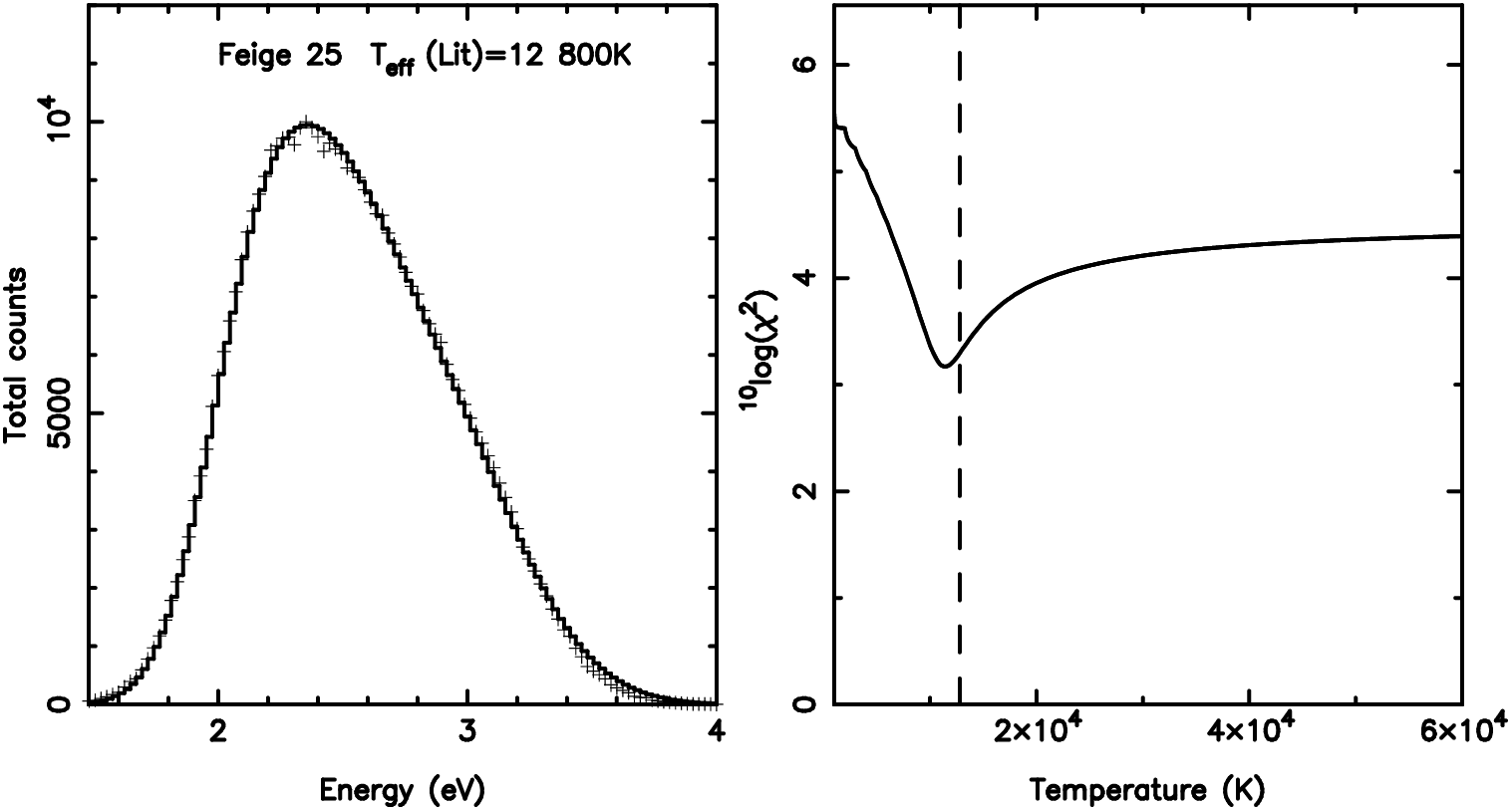}
\caption{Results for EV Lac, G138-31 and Feige 25. Left: the model
fits to the PHA data (data points indicated by crosses; model curves
by histograms). The fits are shown against energy rather than
channel number for clarity.
Right: the dependence of the fit statistic $\chi$$^{2}$ (plotted 
logarithmically)
with
temperature $T$. 
Vertical dashed lines indicate the literature values of $T_{\rm
eff}$.}
\label{Figfits}
\end{figure}

\subsection{The Instrument}

S-Cam2 is ESA's second STJ camera and was used on the three campaigns
during which the observations were made. The instrument was mounted at
the Nasmyth focus of the WHT during all the three campaigns. The
camera contains a 6 $\times$ 6 staggered array of 25 $\times$ 25
$\mu$m$^{2}$ (0.6 $\times$ 0.6 arcsec$^{2}$) tantalum junctions. The
arrival times of individual photons are recorded with an accuracy of 5
$\mu$s, with an energy resolving power $R$ $\sim$ 8. The wavelength
response of the instrument is 310--720 nm. The violet cutoff is
determined by the atmospheric absorption window, while the red cutoff
is caused by the presence of long-wavelength filters designed to
reduce thermal noise. Observations were made in conditions ranging
from good to poor seeing, and with airmasses in the range 1.05--1.54.

\section{Data reduction}

Although we are dealing with optical spectra, our techniques and tools
are those of high-energy astrophysics and we therefore find it more
natural to discuss our data in the energy domain. We are typically
dealing with photons in the range $\sim$ 1.5---4~eV, about a thousand
times less energetic than soft X-ray energies.

The output from a single S-Cam2 observation consists of a list of
photon arrival times, together with the pixel in which each event was
recorded and the channel in which the event has been assigned by the
pulse height analyser (PHA) electronics. The assigned channel depends
linearly on the charge released by each photon when it arrived, which
in turn depends linearly on the energy of the photon. The overall
relationship between photon energy and channel number is therefore
given by $E_i = G \cdot E_p + C$, where $E_p$ is the energy of the
incident photon in eV, and where $G$ and $C$ are the gain (in channels/eV)
and offset (in channels) of the relevant pixel. The gain and offset
properties of all pixels have been mapped and found to be linear and
stable for a given temperature of the detector holder. However small
variations in the detector holder temperature and the electronic
environment of the array can cause variations in gain of $\leq$ 1 \%
during the course of a set of observations. This gain `jitter' cannot
be determined independently of the observations and therefore sets a
limit on the accuracy with which temperatures can presently be
determined.

Data reduction proceeds as follows. We first select the time interval
of data which is to be used to generate an output spectrum. If
necessary, spatial filtering may also be applied at this stage by
selecting and deselecting pixels, e.g., utilising only the pixels
which see sky rather than source. For each selected pixel we then
adjust the counts in each PHA channel to remove known
analogue-to-digital artifacts. This results in a PHA spectrum at the
particular gain and offset of that pixel. We then resample this
spectrum such that it is on a common gain scale corresponding to a
well-behaved reference pixel, and then co-add the contributions from
all selected pixels, making an appropriate correction for the
small differences in detection efficiency from pixel to pixel.
The gain and offset of the reference pixel are
assumed to be $G = 42.5$~channels~eV$^{-1}$, and $C = -2.0$~channels.

The extracted spectra will also contain a contribution from the sky
background which --- depending on the source brightness --- may need
to be removed before model fitting can take place. The sky
contribution can be estimated in two ways. If the seeing was good,
then the corner pixels of the array may be used to obtain a sky
spectrum. The advantage of this approach is that it ensures that the
sky brightness is appropriate for the observation in question. The
second approach is to use a sky frame obtained close (in time and
position) to the observation in question. This approach has the
advantage that many pixels are being sampled for the sky, smoothing
out variations in gain away from the nominal values. In the case of
this study we used corner pixels when the seeing was good, but when
this was not the case we relied on sky frames taken as close as
possible to the observation. These were not dedicated sky frames, but
the results are relatively insensitive to the details of the
background subtraction; indeed for most of the objects in our sample
the background contribution is small enough to have little effect on
the measured temperature.

\section{Temperature Determinations}

\subsection{Method}

We discuss two complimentary approaches to temperature
determination. In both cases we compare model spectra with the
observed distribution of photon energies, searching for the closest
match. The first approach uses synthetic black body spectra 
(Sect.\ref{subsec:BB}), while the second uses a library of stellar templates
of varying spectral type (Sect.\ref{subsec:ST}).

In both cases due allowance is made for the finite energy resolution
of S-Cam2, the transmissive properties of the atmosphere, telescope
optics and detector elements, and the overall quantum efficiency of
the detector.  We make no allowance
for the reddening effects of circumstellar and/or interstellar
extinction, although our results suggest that this
is not a serious drawback, at least in relation to our current 
sample
of stars. For the model fitting, use is made 
of the XSPEC spectral analysis package
(Arnaud 1996), a well-known tool for X-ray astronomy, but
which is equally suited to the analysis of low-energy-resolution
optical data. Apart from the data (including target and background
files), the two inputs that XSPEC requires are a response matrix and a
model spectrum with one or more adjustable parameters. The response
matrix specifies the transformation of the model spectrum from
infinite energy resolution to the degraded energy resolution of the
device, with the folding in of the overall detector efficiency. XSPEC
then finds the set of model parameters that result in the lowest value
of $\chi$$^{2}$. The package contains a large number of internal
models (including a black body) but these were designed for the
analysis of satellite data and are not modified for the effects of
atmospheric absorption. We therefore supply our own models, adjusted
for the mean atmospheric transmission corresponding to the specific
airmass of each stellar observation. We created a grid of models
spanning temperatures between 1000 and 100\,000~K, and then used XSPEC
to find the temperature corresponding to the best-fit black body,
interpolating between grid values.

 \begin{figure}
\centering
\includegraphics[angle=-90,width=8.43truecm]{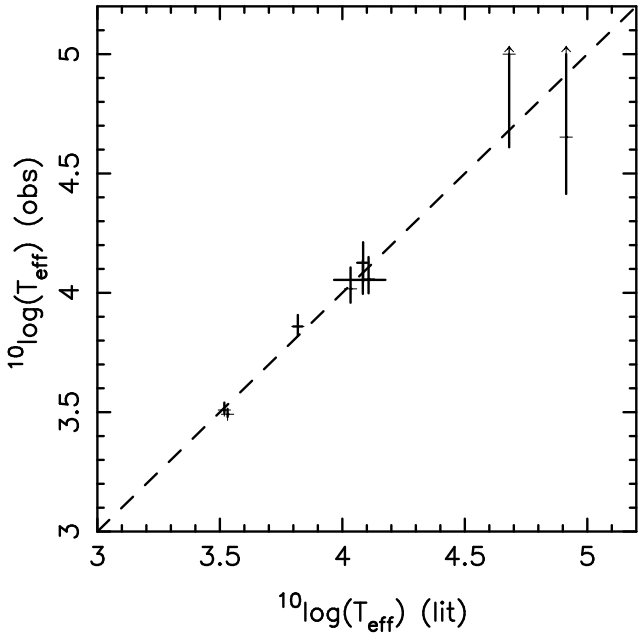}
\caption{Observed $T_{\rm eff}$ versus literature temperatures for the
nine stellar objects in our sample. The vertical error bars indicate
the extreme range of temperatures corresponding to variations in the
gain by $\pm$ 1 $\%$. The dashed line shows the nominal 1:1
correlation.}
\label{Figmainplot}
\end{figure}

\subsection{Results: black body fits}
\label{subsec:BB}

In Fig.~1 we show the data and model fits for three of our
observations, together with the corresponding curves of temperature
versus $\chi$$^{2}$. Data channels 61--179 are included in
the fits, corresponding to  the $\lambda
\sim 310$--$850$~nm bandpass of the instrument, plus an additional
10 channels at the red and blue ends to account for the overspill
of detected events due to the finite energy
resolution.

The models are the input black bodies after modification for
atmospheric transmission and convolution with the overall detector
response. As is clear from the minima in the adjacent $\chi^{2}$
versus $T$ plots, the fits are not formally acceptable, for which one
would expect $\chi^{2}$ in the range 50--150. Nonetheless there are
well-defined minima in the $\chi^{2}$ versus $T$ curves, indicating
the validity of the fitting process. These are all stars with
literature $T_{\rm eff}$ $<$ 15\,000~K. The two hottest stars in our
sample both have $T_{\rm eff} > 40\,000$~K, and in these cases there
is no minimum in the $\chi^{2}$ versus $T$ curve, since for these
objects we are only fitting the red tail of the black body curve,
which varies only slowly for large increases in temperature.

In Fig.~2 we show the observed versus literature temperatures for
the nine stars in our sample. Horizontal error bars on some points
indicate published uncertainty in the literature temperatures. Where
there are two or more literature values we use a mean value (see Table~1
for notes on literature temperatures and adopted means).
 The
vertical error bars reflect the uncertainty in the gain at the time of
the observation, which is typically much larger than the uncertainty
introduced solely by the fitting process (but see Sect.5). We
estimate the gain uncertainty in the following fashion.

It is known that small (mK) variations in the temperature of the
detector holder in the S-Cam2 cryogenic system lead to short-term
variations in the gain on a timescale of minutes to hours, but the
instantaneous gain can presently only be measured directly when
observations are not in progress. This is done using an internal
calibration source consisting of a red LED which illuminates the whole
array. Several LED frames were obtained most nights, with $\sim$
10--15 frames per campaign. An analysis of the LED data showed that
the gain varied from one LED frame to another with a standard
deviation (1 $\sigma$) of 0.4 channels eV$^{-1}$, or $\sim$ 1 \% of
the gain. If this gain `jitter' is also applicable to the observation
data set, we should expect that $\sim$ 67 \%, or $\sim$ 6, of our
observations, should have gains that lie within 0.4 channels eV$^{-1}$
of the nominal gain of 42.5 channels eV$^{-1}$. All else being
correct, therefore, $\sim$ 6 of our observations should have error
bars that intersect the diagonal relationship between $T_{\rm eff}$
(literature) and $T_{\rm eff}$ (observed), with $\sim$ 2--3
outliers. To investigate whether or not this is the case, we therefore
regenerated the response matrix for $\pm$ 1.0 $\sigma$ extreme gain
ranges and remeasured the data using the black body fits. The
resultant temperatures for the upper and lower gain cases are
indicated by the vertical error bars. As is evident from Fig.~2,
seven targets have error bars that intersect the diagonal, with only
two outliers, one of which has rather an uncertain literature
temperature. We therefore conclude that the only significant source of
error in our analysis is due to the uncertain gain variations, and
that this is entirely consistent with the gain jitter as characterised
by the LED frames. Note that for the two hot stars in our sample, the
black body fits reach the 100\,000~K model limit at the central gain
and/or the lower gain limit, and so the error bars are only
constrained at the lower temperature range.

\subsection{Results: stellar template models}
\label{subsec:ST}

The results shown in Fig.~2 suggest that the method works at least
as well for cool stars (M-types) as it does for the hotter objects,
and that there is no significant source of error beyond the gain
uncertainty. Nonetheless, while spectra of hot white dwarfs do indeed
resemble black bodies, this is not the case for the line-dominated
cooler stars. To investigate whether the use of realistic spectra
would improve the determinations, we made use of a library of digital
spectra obtained at Kitt Peak National Observatory (Jacoby et al. 
1984). The library contains 161 stars with spectral types
between M and O. We selected 20 stars of luminosity class V, with
spectral types between M5V and 05V. The digital spectra were then used
as input models in the spectral fitting process, again folding in
atmospheric absorption. In this case we simply selected the template
that gave the lowest $\chi^{2}$ against an observation.

The resultant minimum determines the spectral type of the observation,
from which an estimate of the temperature can be made. This is of
course a less direct determination of temperature than black body
fitting. Given the relatively small number of objects in our template
selection, we estimate that the precision of the spectral class
assignment is $\simeq$ 2 subclasses. Here we discuss only the
non-degenerate stars in our sample. AD Leo and EV Lac are both cool M3.5
flare stars, while Feige 15 is an A0 star. 
For AD Leo and EV Lac we determine the spectral type to be 
M5 in both cases, slightly cooler than the true
spectral type of M3.5. For Feige 15 we find a spectral type of B6,
somewhat hotter than the true value of A0. The results are therefore
acceptable within the anticipated error, but show no obvious
improvement in comparison with the black body estimates. Improved
results might be obtained with a larger set of templates, but
for the time being there is no clear disadvantage in using the
pure black body models, even for cooler stars.

\subsection{Spectral distortion due to pile-up}

In common with all photon counting detectors, S-Cam2 has a finite
response time associated with the arrival of a single photon. If a
second photon arrives during this interval, the detection of one or
both may be affected. Although a naive consideration might suggest
that two photons arriving within a short time would always lead to a
higher energy being detected, in practice the detailed properties of
the electronic shaping mean that possible outcomes can include not
only the assigning of higher energies to one or more events, but also
the assigning of lower energies or even the non-detection of an event.
The net result is that as count rates increase, recorded spectra
suffer increasing distortion, with events being redistributed to both
higher and lower energy channels.

The electronics of S-Cam2 are such that only photons arriving on a
single pixel within $\sim$ 200 $\mu$s of a preceding photon will be
affected by pile-up. Pile-up should therefore be of little consequence
for count rates below 500--1000 Hz per pixel, a prediction which has
been verified in the laboratory. Since the maximum count rates of our
stellar observations are all $<$ 800 Hz per pixel, pile-up is not
expected to cause significant error in the derived temperatures. To
explore this further, we have simulated the electronic response of
S-Cam2 to the rapidly varying intensity profile of the Crab pulsar, an
object that was observed during several S-Cam campaigns (Perryman et
al. 1999). The instantaneous intensity of the Crab pulsar and its
background nebula component ranges from $\sim$ 100 counts~s$^{-1}$ per
pixel in the interpulse to $>$ 2000 counts~s$^{-1}$ per pixel during
the main pulse. Our phase-resolved observations of the Crab were split
into three energy bands, corresponding roughly to red ($R$), visual
($V$) and blue ($B$). From these three energy bands we construct two
colour ratios: $V/R$ and $B/V$, and plot them against phase
(Fig.~3). Significant variations are apparent, especially in the
$B/V$ data, but after dividing out the predicted variations due to
pile-up, there is no longer any strong evidence for colour changes
across the pulse profile. This is consistent with previous studies in
the UV-optical-IR band, which have reported either no colour changes
(e.g., Carrami\~{n}ana et al. 2000), or very slight ones (e.g.,
Eikenberry et al. 1996, Romani et al. 2001). This suggests that the
pile-up characteristics of S-Cam2 are well understood, at least in the
count rate regime probed by the Crab.

\begin{figure}
\centering
\includegraphics[angle=0,width=8.43truecm]{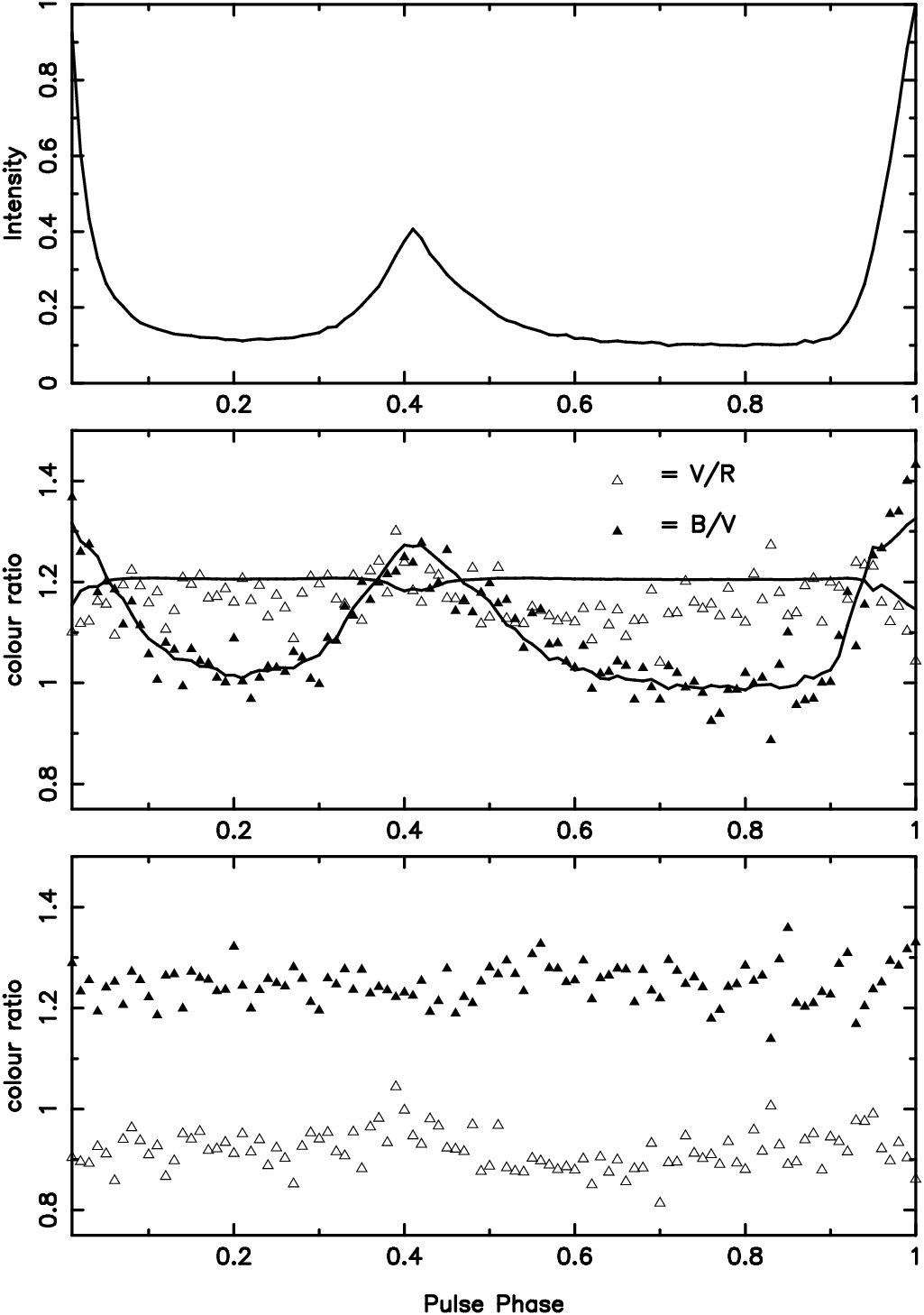}
\caption{Colour ratios for the optical light curve of the Crab
pulsar. Top: the observed intensity profile of the Crab. Middle: the
measured $V/R$ and $B/V$ colour ratios (symbols) and the predicted
variations for the same ratios (lines). Bottom: the measured colour
ratios after dividing by the predicted variations. Arbitrary offsets
have been applied for the sake of clarity.}
\label{Fighardnessratios}
\end{figure}

Using the same approach, we have created a synthetic input stellar
spectrum and subjected this to simulated pile-up distortion for count
rates ranging from 10--3000 counts~s$^{-1}$ per pixel. At each count
rate we determine the temperature via the same black body fitting
method outlined earlier. We generated ten simulations at each count
rate step, measuring the temperatures individually and deriving their
mean and variance.

The results, shown in Fig.~4, indicate that pile-up has no
significant effect on the measured temperature for count rates below
$\sim$ 250 counts~s$^{-1}$ per pixel, and is modest for count rates
$\leq$ 750 counts~s$^{-1}$ per pixel. Once the count rate exceeds
$\sim$ 1000 counts~s$^{-1}$, however, the distortion becomes severe
and pile-up rivals gain uncertainty as the main source of error in
temperature determination. Similar conclusions are drawn using a
hotter input spectrum. None of the stellar observations discussed in
this paper had individual pixel count rates in excess of 700
counts~s$^{-1}$, indicating that pile-up is not the major source of
error in our data. In future versions of the instrument, however,
where the gain variations are likely to be better constrained, an
accurate treatment of pile-up may be necessary in order to extract the
most accurate temperature determinations.

\begin{figure}
\centering
\includegraphics[angle=0,width=8.43truecm]{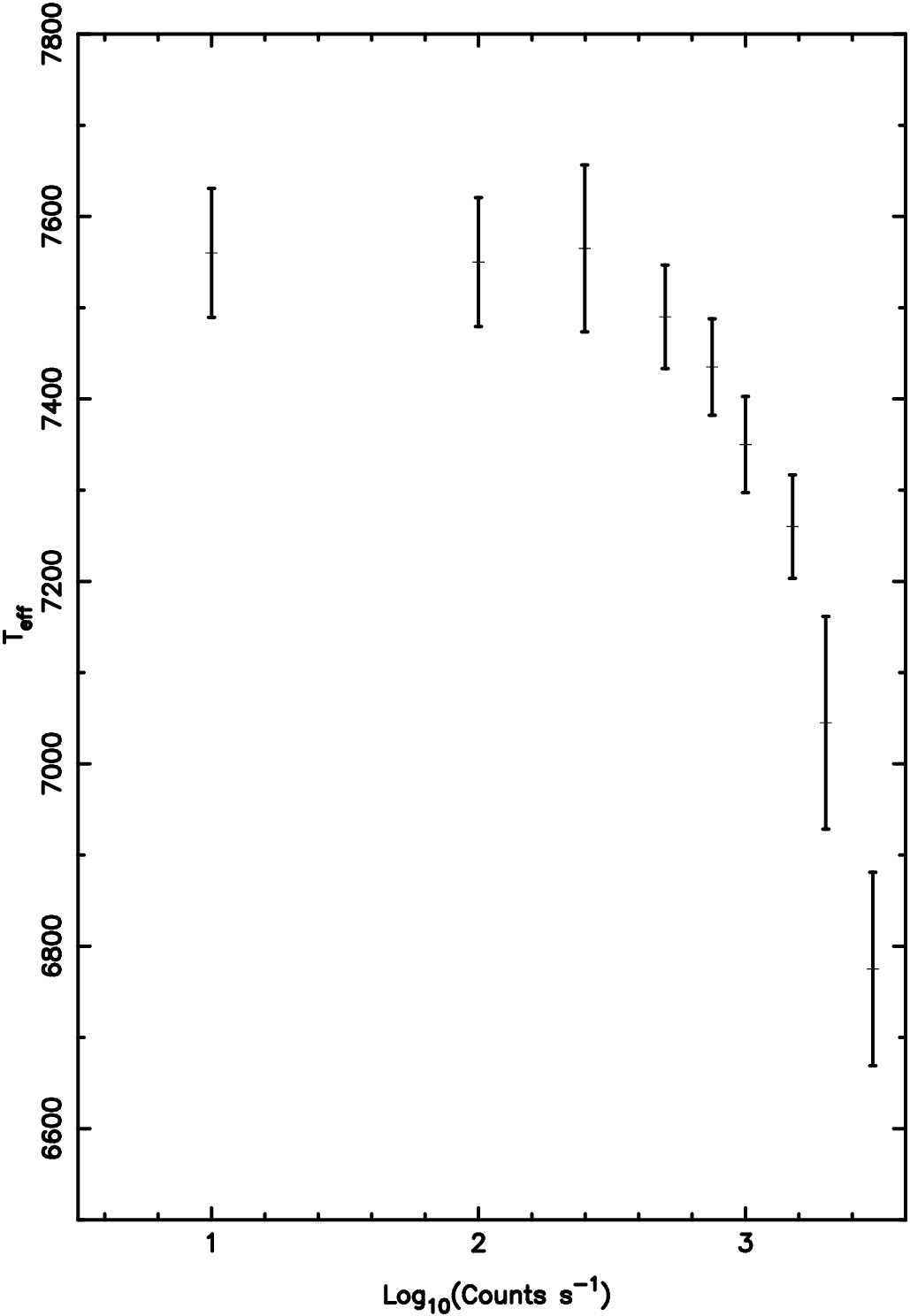}
\caption{The predicted variation in measured temperature with count
rate, due to distortion of the spectral profile by pile-up. Each data
point is derived from the mean of 10 simulations, with 1 $\sigma$
error bars. Note that the measured temperature shows no significant
drop until count rates exceed $\sim$ 500 Hz.}
\label{Figtemp_countrate}
\end{figure}

\section{Application of the technique: cataclysmic variable stars}

During the course of the three S-Cam2 campaigns we also made a number
of observations of eclipsing cataclysmic
variables (CVs). Three systems were polars and one a dwarf nova.
Reports on the S-Cam2 observations of the polars
UZ For, HU Aqr and EP Dra may be found in Perryman et
al. (2001), and Bridge et al. (2002a, 2002b). S-Cam2 observations of
the dwarf nova IY UMa are reported in Steeghs et al. (2002).

The main
difference between these systems is in the accretion process,
which is in turn determined by the strength of the magnetic
field. In polars the field is sufficiently strong that
accretion stream from the secondary becomes 
magnetically confined in the vicinity of the dwarf. From this point 
accretion material follows magnetic field lines to
 the dwarf surface, where accretion occurs through a stand-off shock. 
In the dwarf novae, the magnetic field is weak or absent, such that
accretion is via a disk.

\begin{figure}
\centering
\includegraphics[angle=-90,width=8.43truecm]{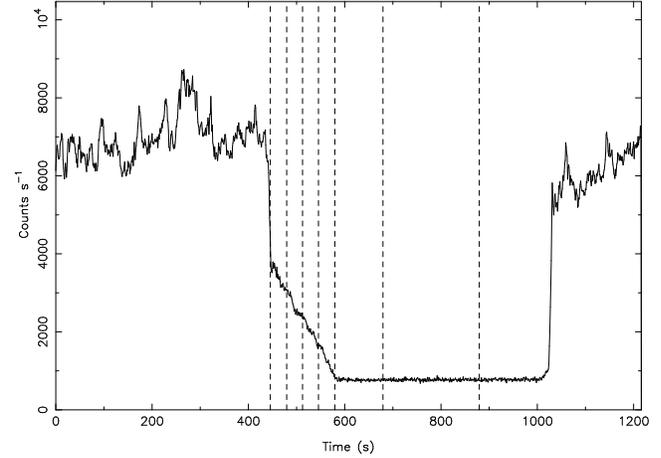}
\caption{An eclipse light curve of the polar CV HU Aqr. The
vertical bars indicate the eclipse of the accretion stream (divided
into four 33 s components) and the 200 s portion of the mid-eclipse
used for background subtraction.}
\label{Fighuaqr}
\end{figure}

Prior to the eclipse, the dominant source of optical emission is the
central accretion region, be it the white dwarf, one or more hotspots
on the surface, and/or a bright accretion column. Once this region is
occulted by the donor star, however, cooler, fainter components become
amenable to study. In the polars these include the rear
hemisphere of the donor and the accretion stream itself. In dwarf
novae, the disk/stream interaction is seen in addition to the
contribution from the secondary. In principle, the stream or disk
temperatures can be measured provided the contribution from the
secondary is small enough to ignore, or can be subtracted. S-Cam2 is
obviously well suited to this kind of study, since the event list data
allow spectra to be accumulated across any specified time interval.

\subsection{Stream temperatures in magnetic CVs: HU Aqr, EP Dra, UZ For}

HU Aqr is a polar with a period of $\sim$ 125
min. S-Cam2 observations of this binary are discussed in detail in
Bridge et al. (2002b). Three eclipses were observed, each of which was
characterised by a very sharp ingress of the bright accretion region,
followed by a more gradual ingress of a component identified with the
accretion stream. The eclipse with the best S/N is shown in Fig.~5,
where the accretion stream ingress is very well defined. By studying
hardness ratios, Bridge et al. concluded that the light from the
system became significantly redder once the accretion region was
eclipsed, as expected given that the stream is likely to be much
cooler than the accretion region. However the error bars on the data
were too large to put any strong constraints on the temperature
gradient along the stream.

\begin{figure}
\centering
\includegraphics[angle=-90,width=8.43truecm]{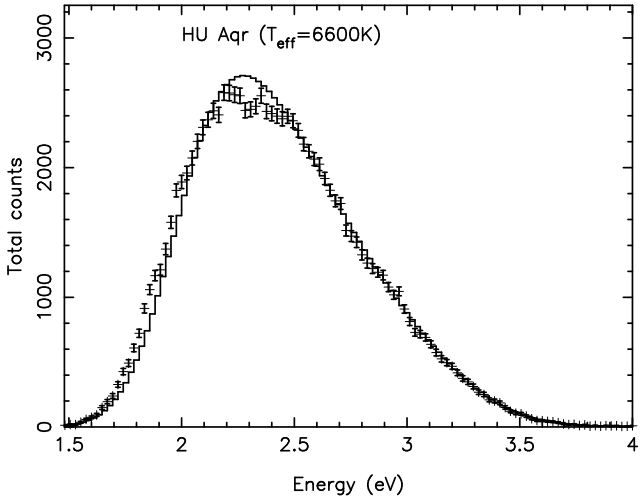}
\caption{The black body model fit to the first 33 s portion of the
stream data in HU Aqr. The data has been background subtracted.}
\label{Fighuaqrfit}
\end{figure}

\begin{figure}
\centering
\includegraphics[angle=-90,width=8.43truecm]{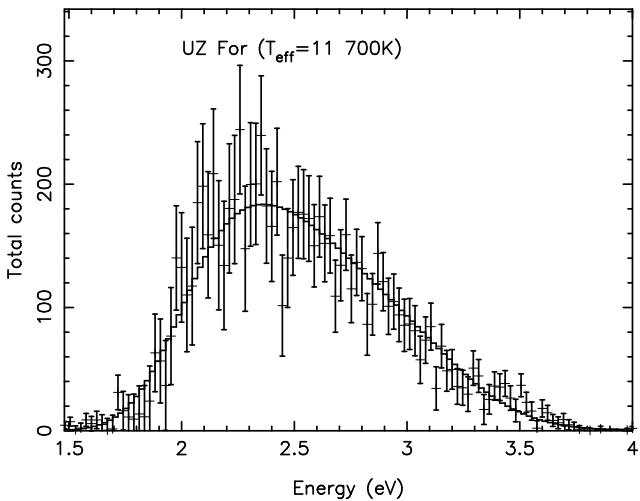}
\caption{The black body model fit to the stream component in UZ
For. The data has been background subtracted.}
\label{Figuzforfit}
\end{figure}

We have now reexamined the HU Aqr data, this time using the
temperature fitting technique. In all three eclipses, a spectrum
corresponding to the stream was isolated, together with a portion of
mid-eclipse data to use for background subtraction. Using the
mid-eclipse as a background has the advantage that we are also
subtracting any residual light from the donor star, thereby isolating
only the stream.

Of the three eclipses observed by S-Cam2, the first two had
insufficient S/N (due to seeing, and/or weakness of the feature) to
enable the stream to be split into sub-components. The first eclipse
yielded $T_{\rm eff} = 6700 \pm 600$~K, while the second yielded
$T_{\rm eff} = 6500 \pm 500$~K. The third eclipse could be split into
four time-resolved components, as indicated in Fig.~5. These gave
--- in increasing time order --- $T_{\rm eff} = 6600, 6600, 6700$ and
$6300$~K, with an estimated error on each measurement of $\pm$ 600
K. The fit to the first segment of stream data is shown in Fig.~6.
Clearly there is no evidence for a gradient in temperature along the
stream. Using $UBVR$ photometry, Harrop-Allin et al. (1999) estimated
the stream temperature in HU Aqr to lie in the range 6500--7400~K,
with 6500~K being their preferred estimate, in satisfactory agreement
with the results reported here. We thus have confidence in extending
the technique to other systems.

UZ For and EP Dra are also eclipsing polars which have been the
subject of S-Cam2 observations (Perryman et al. 2001, Bridge et
al. 2002a). In both cases a stream is sometimes seen, though it is
not as well-defined as in HU Aqr. We used the same approach as above,
isolating stream/eclipse regions. In the case of EP Dra we observed
clear stream ingresses in three eclipses, yielding $T_{\rm eff}$
values of 5700~K, 7200~K and 6050~K. The uncertainties were also
$\sim$ $\pm$ 600~K. In this case there is a possible source of
systematic error since certain pixels were somewhat unstable during
the observations. Clearly, however, the fits point to the stream
having a temperature in the same range as that seen in HU Aqr.

In the case of UZ For there was only one eclipse for which the data
quality was sufficiently high to permit the extraction of a stream
spectrum. Here, $T_{\rm eff}$ was found to be 11\,700~K $\pm$ 2500~K,
significantly higher than for the other two CVs. The fit is shown in
Fig.~7. The quoted error includes the uncertainty associated with
the model fit, in addition to the uncertain gain (normally our
statistics are such that the gain uncertainty is the only significant
source of error).

Such a high temperature is not necessarily unreasonable, since the
observed temperature of the stream is influenced by a number of
processes. The most important are irradiation by X-rays from the white
dwarf and magnetic heating in the threading region where material is
magnetically confined. The amount of magnetic heating will be related
to the amount and density of material in the stream, and is possibly
caused by magnetic reconnection or turbulence (Li 1999) or small
shocks (Liebert \& Stockman 1985; Hameury et al. 1986). The
amount of irradiative heating of the stream will be dependent upon the
area of the stream presented to the white dwarf. The temperature that
we determine is then influenced by the area of this heated stream face
presented to the observer.

The larger temperature found for UZ For is therefore probably a
consequence of observing a larger area of the heated stream face, and
stream material that is heated more in the threading region. The UZ
For S-Cam2 observations show evidence for two accretion regions (see
Perryman et al. 2001). This means that the field lines along which
material accretes are filled above and below the orbital plane. We
therefore observe more of the heated face of the stream than in HU Aqr
or EP Dra where there is evidence for accretion at only one region on
the white dwarf (Bridge et al. 2002a, 2002b). Material is still
probably heated in HU Aqr at the threading region, as found by
Vrielmann \& Schwope (2001) and Harrop-Allin (1999) previously, and
this also probably occurs in EP Dra. However, the viewing angle to the
stream is such that we do not see as large an area of the heated
stream face. Future observations of UZ For will aim to improve the S/N
of the stream component.

\begin{figure}
\centering
\includegraphics[angle=-90,width=8.43truecm]{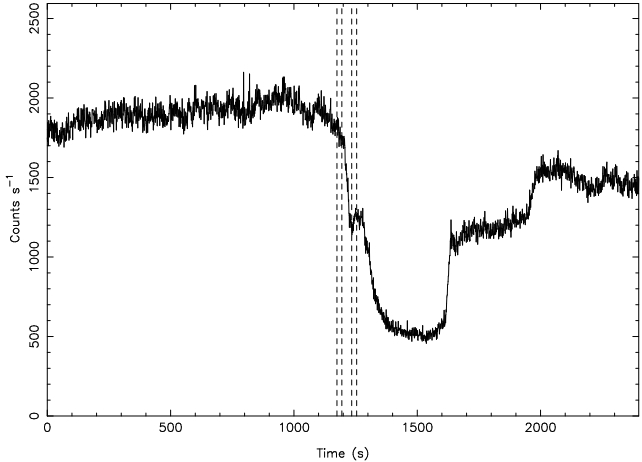}
\caption{An eclipse light curve of IY UMa, a non-magnetic CV of the 
dwarf novae type. The
vertical bars indicate the two 20 s segments of the eclipse used to
isolate the spectrum of the white dwarf, immediately before and
immediately after the white dwarf ingress.}
\label{Figiyuma}
\end{figure}

\subsection{The temperature of the white dwarf in the non-magnetic CV IY UMa}

IY UMa is a non-magnetic CV in which accretion proceeds via a disk.
The S-Cam2 observations are discussed in Steeghs et al. (2002). The
eclipse morphology is significantly more complex (see Fig.~8, and
the discussion in Steeghs et al.). In this case, the geometry
precludes easy selection of an accreting region characterised by a
single temperature. However it is possible to isolate the spectrum of
the white dwarf by extracting the difference between the spectrum of
the source immediately before and immediately after the very rapid
eclipse of the compact component. Three eclipses were observed by
S-Cam2, but the low S/N requires the summing of these extracted
spectra prior to model fitting. Using a co-added spectrum, therefore,
we obtained $T_{\rm eff} = 16\,000 \pm 3000$~K for the white dwarf in
IY UMa (Fig.~9). As in the case of UZ For, the quoted error includes
the uncertainty due to the fitting process.

G\"{a}nsicke \& Koester (1999) list the effective temperatures for the
white dwarfs in the 9 dwarf novae for which the white dwarf spectrum
had been unequivocally identified in the UV or optical at the time of
their paper. Their list does not include IY UMa. Excluding the
long-period system U Gem, which lies above the CV period gap and is
significantly hotter than those short-period systems below it, the
results span the temperature range 14\,000--22\,000~K, with a mean
value of $\sim$ 17\,000~K. The value we have obtained for IY UMa is
therefore quite plausible. Sion (1999) presents a more extensive list
of non-magnetic white dwarf temperatures, of which 9 are dwarf novae
of the same SU UMa type as IY UMa. Of the 9 objects in Sion's sample,
7 are also found in the G\"{a}nsicke \& Koester list, but in most
cases the quoted temperatures are somewhat different. The
temperatures of the Sion sample range from 12\,000--18\,500~K, with a
mean value of 15\,600~K. Our value is again consistent with
expectations.

\begin{figure}
\centering
\includegraphics[angle=-90,width=8.43truecm]{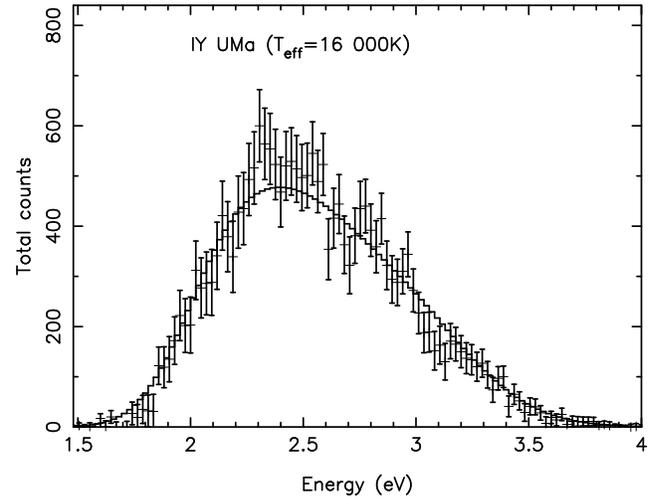}
\caption{The black body model fit to the summed spectrum of the white
dwarf component in IY UMa. See text for details.}
\label{Figiyumafit}
\end{figure}

\section{Conclusions and future prospects}

In this paper we have outlined a fundamentally new approach to
temperature determination using non-dispersive optical
spectroscopy. The spectra are modelled with black bodies, modified for
atmospheric absorption. For cool stars ($T_{\rm eff} < 15\,000$~K),
the derived temperatures are generally 
within 10\% of the literature values. The
agreement is less good for hotter stars, but the main sources of error
are calibration uncertainties, rather than any intrinsic limitation
with the method.

Having demonstrated that the technique works, we have applied it to
four cataclysmic binary stars, isolating accretion stream components
in three cases and the white dwarf itself in a fourth. The time-tagged
photon event list data format of S-Cam2 data makes the isolation of
these components relatively straightforward. Where appropriate, the
results are compared against literature values and found to be in
satisfactory agreement.

Work is now underway on the development of S-Cam3, which will
incorporate $10 \times 12$ pixels, each of which is slightly larger
than those used in S-Cam2. The field of view will therefore be much
improved, allowing the edges of the frame to be used for background
subtraction with no fear of contamination by source counts. In
addition, the energy resolving power is likely to be improved by
20--30\%, and realtime monitoring of the gain variations will be
possible. All these factors should lead to a significant improvement
in the quality of the temperature determination possible with the
instrument, in addition to other model-fitting problems such
as quasar redshift determination.

\begin{acknowledgements}

We thank the other members of the S-Cam2 team for their
enthusiastic support throughout the program. In addition,
Erik Kuulkers and Danny Steeghs are thanked for useful discussions
regarding the CV temperature fitting. The William Herschel Telescope
is operated on the island of La Palma by the Isaac Newton Group in
the Spanish Observatoria del Roque de los Muchachos of the Instituto
de Astrofisica de Canarias. We thank the anonymous referee for
useful comments on this manuscript.

\end{acknowledgements}

\end{document}